\documentclass{article}
\usepackage{arxiv}

\usepackage{subfigure}

\usepackage[utf8]{inputenc}
\usepackage{algorithm} %ctan.org\pkg\algorithms
\usepackage{algpseudocode}
\usepackage{tikz}
\usetikzlibrary{backgrounds,fit,decorations.pathreplacing, shapes}  % TikZ libraries
\usepackage{tkz-graph}

\usepackage[super]{nth}
\usepackage{adjustbox}
\usepackage{nameref}

\tikzstyle{vertex}=[circle,draw=black, fill=white,sloped,minimum size=17pt,inner sep=5pt]

\usepackage{tikz}
\usetikzlibrary{decorations.pathreplacing,calc}

\usepackage{enumerate}
\usepackage{amsmath}
\usepackage{graphicx}
\usepackage{wrapfig}
\usepackage{lscape}
\usepackage{rotating}
\usepackage{epstopdf}

\usepackage{caption}
\usepackage{url}
\usepackage{hyperref}
\usepackage{float}
\usepackage[super]{nth}
\usepackage[labelfont=bf]{caption}
\usepackage[labelsep=period]{caption}
\usepackage{pgfplots} % LaTeX

\pgfplotsset{compat = newest}

\newcommand{\figref}[1]%
{Figure \ref{#1}%
}

\newcommand{\tableref}[1]%
{Table \ref{#1}%
}

\newcommand{\algorithmref}[1]%
{Algorithm \ref{#1}%
}

\newcommand{\sectionref}[1]%
{Section \ref{#1}%
}

\newcommand{\lineref}[1]%
{Line \ref{#1}%
}

\algnewcommand{\LineComment}[1]{\State \(\triangleright\) #1}

\usepackage{newunicodechar}
\newunicodechar{−}{-}

\title{Incremental Sparse TFIDF \& Incremental Similarity with Bipartite Graphs}

\author{Rui Portocarrero Sarmento \\
LIAAD-INESC TEC \\
PRODEI - Faculty of Engineering, University of Porto \\
mail@ruisarmento.com \And
Pavel Brazdil \\
LIAAD-INESC TEC \\
pbrazdil@inesctec.pt
}

\begin{document}

\long\def\/*#1*/{}

\maketitle

\begin{abstract}
    In this report, we present several experiments with TF-IDF and associated cosine similarity
\end{abstract}

\keywords{Incremental Sparse TFIDF \and Data Streams \and Text Streams \and Incremental Similarity.}

\section{Introduction}

In this report, we experimented with several concepts regarding text streams analysis.

We tested an implementation of Incremental Sparse TF-IDF (IS-TFIDF) and Incremental Cosine Similarity (ICS) with the use of bipartite graphs.

We are using bipartite graphs - one type of node are documents, and the other type of nodes are words - to know what documents are affected with a word arrival at the stream (the neighbors of the word in the graph). Thus, with this information, we leverage optimized algorithms used for graph-based applications. The concept is similar to, for example, the use of hash tables or other computer science concepts used for fast access to information in memory.

\section{Tools}

For this report we used several R packages. To compute graph-based operations we used the "igraph" package \citep{ww}. Text mining and preprocessing was done with the "tm" package \citep{w1,w2}. "Rcpp" package was used to provide libraries for packages built with C++ \citep{x,x1,x2}. To take measurements of processing and memory use we used package "microbenchmark" \citep{z}.

\section{IS-TFIDF for Text Streams}\label{Inc}

In this section, we explain the method and our settings for the tests of these two versions of IS-TFIDF.

\subsection{IS-TFIDF and ICS Method}

IS-TFIDF is an updatable list structure of documents. Each entry per document has a vector of words that the document has, after typical pre-processing (removal of stop words, numbers, etc.). This vector contains the TF-IDF values for each word. These values are also updated in each iteration of the stream.

In each new iteration or input of a new chunk of text documents, we update a bipartite graph by adding new words and documents and establishing the connection (edges) between words and documents. Thus, this graph has two types of nodes, words, and documents. New words that appear in the stream are connected to their documents. 

Regarding our goal, the efficient update of the similarity between documents (ICS), we use the bipartite graph first order neighbors for new or updated words in the stream, to check which pairs of documents' changes similarity. We then recalculate these pairs similarity, due to changes. This way, we avoid the recalculation of the similarity of all pairs of documents in the stream, in each iteration of the stream.

\begin{figure}[H]
  \centering
    \includegraphics[scale=0.4]{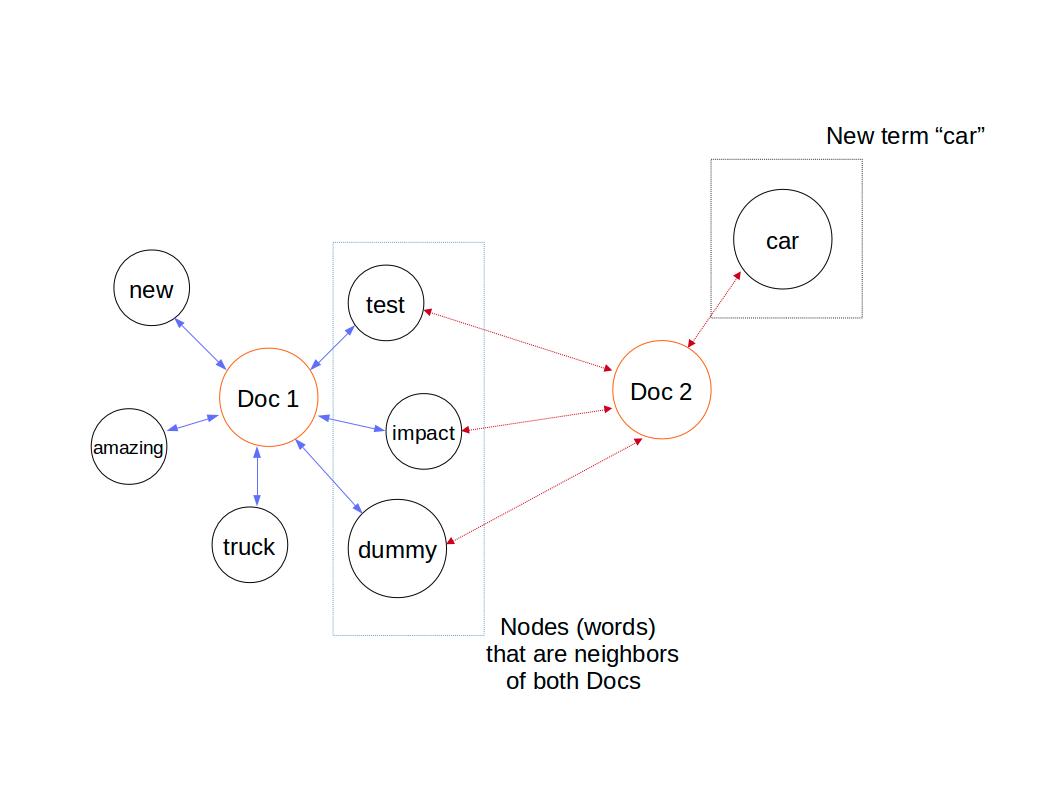}
    \caption{Bipartite Graph Method}
\label{fig:BipMethod}
\end{figure}

~\figref{fig:BipMethod} shows an example of our method. This example has two snapshots or iterations of a stream. Please consider this example as a sample of two documents in a larger network of documents and words. 

\begin{itemize}
    \item First Snapshot of text (Doc 1): ``New Amazing Truck Impact Test Dummy''
    \item Second Snapshot of text (Doc 2): ``Car Impact Test Dummy''
\end{itemize}

``Car'' is a new term added to the graph with the appearance of Doc 2. Additionally, the words that are simultaneously common to both documents also change the TF-IDF values. Please note that, if Doc 2 only had the word ``Car'' we did not need to update the similarity between Doc 1 and Doc 2 since they were not connected (they were not both neighbors of the term ``Car''). Instead, as we have the neighbor words ``Impact'', ``Test'', and ``Dummy'' changing their TF-IDF values in the second snapshot, we have to recalculate similarity between Doc 1 and Doc 2 in this iteration.

\subsection{One Document Streaming (ODS)}

The first version we experimented IS-TFIDF with was based on the concept of using a sliding window to iterate through the stream of data, where each snapshot is considered a new document. Thus, this way the input is not incremental in a sense that any new chunk of text is not considered an update or increment in an already processed document.

\subsection{Several Documents Streaming (SDS)}

The second version we experimented IS-TFIDF with was based on the concept of using a sliding window to iterate through the stream of data, where each snapshot is considered a flow of several documents. Consider, for example, that in the first snapshot we had text for doc1, doc2, and doc3. Thus, if in the third snapshot we had more information added to doc1, we would incrementally update and proceed with the addition of information to doc1.

\section{Case Study and Evaluation}\label{Met}
   
In this section, we present the method to analyze both ODS and SDS implementations of IS-TFIDF, regarding its comparison with the original batch algorithm. We introduce the reader to the data used for the tests. Finally, we describe the methodology to perform the comparison tests.

%a rever esta secção
\subsection{Description of the Data}

In this case study, for algorithms efficiency measurements, we selected a dataset publicly available for research purposes by \cite{Signal1M2016}. This dataset has a high amount of information including Reuters news and also several articles extracted from Blogs. The data contains articles titles, content, type of publication (which might include news or Blog posts), the source institution and also date and time of publishing. The high quality and high organization of this structured data make it a good source for text mining or NLP tasks.
We selected all Reuters news, from the 1\textsuperscript{st} to the 30\textsuperscript{th} of September 2015. This corresponds to 400 news articles. Regarding the used text, we could choose between the news Titles or the news content to analyze. We choose the news content in all our studies.

\subsection{Methodology}

%a rever esta secção
The ODS IS-TFIDF and ICS implementation and the SDS IS-TFIDF and ICS implementation were evaluated in an incremental setup, for both the developed versions. Additionally, both algorithms were tested and evaluated regarding Processing Efficiency and compared with the batch version. 

\subsubsection{Processing Efficiency}

Results for the ODS IS-TFIDF and ODS ICS were obtained from several increments of text news, compared with the original batch algorithm, using the algorithms in the tm package of R. Both versions were tested regarding efficiency in an incremental setting configuration. The original batch results served as a baseline. In this setup, six snapshots of news were considered. In total, the first 20 days of September 2015, corresponding to 300 news articles, were passed as input. The snapshots were built by aggregating publications on a daily basis, by using the timestamps available in the dataset. In the Batch algorithm, for every snapshot, the input text has all daily news since the day 1 to the current day. For the incremental algorithm (ODS IS-TFIDF and ICS), in the first snapshot, it receives only 15 days of news articles as input. In the following snapshots, the algorithm only receives the set of publications text added to the corpus in that particular day snapshot (incremental).

Results for the SDS IS-TFIDF and SDS ICS were obtained from several increments of INESC TEC researchers data. Each increment had several author's publication titles, with five research titles per each author. In the end, 22 snapshots of data were sent as input.

The empirical evaluation performed consisted mainly of comparing run times of each increment (duration of each increment and cumulative execution time) for all versions of TFIDF and ICS. Note that the batch algorithm will always need to process all the accumulated text. In the end, an analysis of the total speed-up ratio obtained in each of the steps is added.

\section{Results}\label{Res}

We present the results for all versions of the algorithm here in this section. All algorithms are compared with the Batch version of TF-IDF with subsequent cosine similarity calculation. The comparisons regard only the efficiency. 

\subsection{Processing Efficiency}

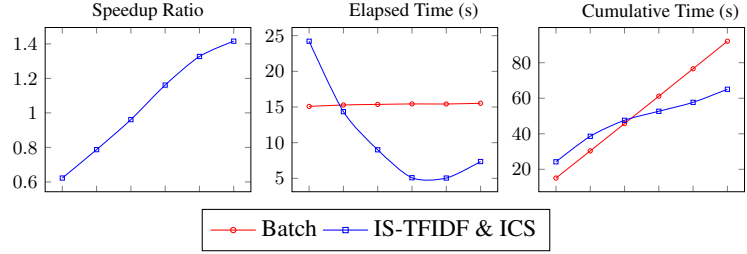
\begin{figure*}
  \centering
    \pgfplotstableread[col sep=semicolon]{images/data/results_Reuters_elapsed_time.txt}\dataTableElapsedTime
\pgfplotstableread[col sep=semicolon]{images/data/results_Reuters_cum_time.txt}\dataTableCumulativeTime
\pgfplotstableread[col sep=semicolon]{images/data/results_Reuters_speedup.txt}\dataTableSpeedup

% Preamble: \pgfplotsset{width=7cm,compat=smooth,mark=*,red,mark size=0.75,mark size=0.53}
\pgfplotsset{footnotesize}
\begin{center}% note that \centering uses less vspace...
%
% Nodes
%

\begin{tikzpicture}[scale=0.8]
\begin{axis}[
title={Speedup Ratio},
%ylabel = nodes,
%ymode=log,
%ymax = 5000,
%ymin = 1000,
xticklabels={,,},
legend style={draw=none}
]
%\addplot[smooth,mark=square,red!50,mark size=0.75] table[y = speedup] from \dataTableSpeedup;
%\addplot[dashed,mark=triangle,blue!50,mark size=0.75] table[y = speedupinc] from \dataTableSpeedup;

\addplot[smooth,mark=square,blue,mark size=1] table[y = speedup] from \dataTableSpeedup;

\end{axis}
\end{tikzpicture}
\begin{tikzpicture}[scale=0.8]
\begin{axis}[
title={\qquad Elapsed Time (s)},
%ymode=log,
%ymax = 50000,
%ymin = 100,
%yticklabels={,,},
xticklabels={,,},
]
%\addplot[smooth,mark=o,green,mark size=0.75] table[y = batch] from \dataTableElapsedTime;
%\addplot[smooth,mark=square,red,mark size=0.75] table[y = Window-based] from \dataTableElapsedTime;
%\addplot[dashed,mark=triangle,blue,mark size=0.75] table[y = Inc] from \dataTableElapsedTime;

\addplot[smooth,mark=o,red,mark size=1] table[y = batch] from \dataTableElapsedTime;
\addplot[smooth,mark=square,blue,mark size=1] table[y = IS-TFIDF & ICS] from \dataTableElapsedTime;

\end{axis}
\end{tikzpicture}
\begin{tikzpicture}[scale=0.8]
\begin{axis}[
legend columns=-1,
legend entries={Batch, IS-TFIDF \& ICS},
legend to name=named,
title={\qquad Cumulative Time (s)},
%ymode=log,
%ymax = 50000,
%ymin = 100,
%yticklabels={,,},
xticklabels={,,},
]
%\addplot[smooth,mark=o,green,mark size=0.75] table[y = batch] from \dataTableCumulativeTime;
%\addplot[smooth,mark=square,red,mark size=0.75] table[y = Window-based] from \dataTableCumulativeTime;
%\addplot[dashed,mark=triangle,blue,mark size=0.75] table[y = Inc] from \dataTableCumulativeTime;

\addplot[smooth,mark=o,red,mark size=1] table[y = batch] from \dataTableCumulativeTime;
\addplot[smooth,mark=square,blue,mark size=1] table[y = IS-TFIDF & ICS] from \dataTableCumulativeTime;

\end{axis}
\end{tikzpicture}
\\
\ref{named}
\end{center}
    \caption{Efficiency results with the Reuters News Dataset (for all algorithms)}
\label{fig:Results1}
\end{figure*}

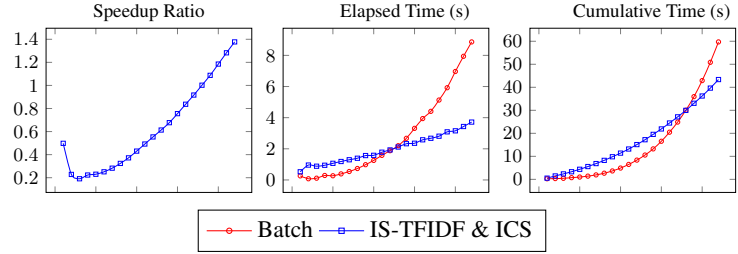
\begin{figure*}
  \centering
    \pgfplotstableread[col sep=semicolon]{images/data/results_Inesc_elapsed_time.txt}\dataTableElapsedTime
\pgfplotstableread[col sep=semicolon]{images/data/results_Inesc_cum_time.txt}\dataTableCumulativeTime
\pgfplotstableread[col sep=semicolon]{images/data/results_Inesc_speedup.txt}\dataTableSpeedup

% Preamble: \pgfplotsset{width=7cm,compat=smooth,mark=*,red,mark size=0.75,mark size=0.53}
\pgfplotsset{footnotesize}
\begin{center}% note that \centering uses less vspace...
\begin{tikzpicture}[scale=0.8]
\begin{axis}[
title={Speedup Ratio},
%ylabel = nodes,
%ymode=log,
%ymax = 5000,
%ymin = 1000,
xticklabels={,,},
legend style={draw=none}
]
%\addplot[smooth,mark=square,red!50,mark size=0.75] table[y = speedup] from \dataTableSpeedup;
%\addplot[dashed,mark=triangle,blue!50,mark size=0.75] table[y = speedupinc] from \dataTableSpeedup;

\addplot[smooth,mark=square,blue,mark size=1] table[y = speedup] from \dataTableSpeedup;

\end{axis}
\end{tikzpicture}
\begin{tikzpicture}[scale=0.8]
\begin{axis}[
title={\qquad Elapsed Time (s)},
%ymode=log,
%ymax = 50000,
%ymin = 100,
%yticklabels={,,},
xticklabels={,,},
]
%\addplot[smooth,mark=o,green,mark size=0.75] table[y = batch] from \dataTableElapsedTime;
%\addplot[smooth,mark=square,red,mark size=0.75] table[y = Window-based] from \dataTableElapsedTime;
%\addplot[dashed,mark=triangle,blue,mark size=0.75] table[y = Inc] from \dataTableElapsedTime;

\addplot[smooth,mark=o,red,mark size=1] table[y = batch] from \dataTableElapsedTime;
\addplot[smooth,mark=square,blue,mark size=1] table[y = IS-TFIDF & ICS ] from \dataTableElapsedTime;

\end{axis}
\end{tikzpicture}
\begin{tikzpicture}[scale=0.8]
\begin{axis}[
legend columns=-1,
legend entries={Batch, IS-TFIDF \& ICS},
legend to name=named,
title={\qquad Cumulative Time (s)},
%ymode=log,
%ymax = 50000,
%ymin = 100,
%yticklabels={,,},
xticklabels={,,},
]
%\addplot[smooth,mark=o,green,mark size=0.75] table[y = batch] from \dataTableCumulativeTime;
%\addplot[smooth,mark=square,red,mark size=0.75] table[y = Window-based] from \dataTableCumulativeTime;
%\addplot[dashed,mark=triangle,blue,mark size=0.75] table[y = Inc] from \dataTableCumulativeTime;

\addplot[smooth,mark=o,red,mark size=1] table[y = batch] from \dataTableCumulativeTime;
\addplot[smooth,mark=square,blue,mark size=1] table[y = IS-TFIDF & ICS] from \dataTableCumulativeTime;

\end{axis}
\end{tikzpicture}
\\
\ref{named}
\end{center}
    \caption{Efficiency results with the INESC TEC research Dataset (for all algorithms)}
\label{fig:Results2}
\end{figure*}

\figref{fig:Results1} (Elapsed Time) shows that the Batch algorithm has a slight increase in elapsed time. Regarding our implementation, the update time decreases as the stream evolves.

\figref{fig:Results1} (Cumulative Time) shows a clear advantage of using IS-TFIDF \& ICS, as the cumulative time rises at a slower pace when compared with the batch version.

\figref{fig:Results1} (Speedup Ratio) shows that, although the IS-TFIDF implementation is slower in the first iterations, it starts being faster after some iterations. 

\figref{fig:Results2} (Elapsed Time) shows that the Batch algorithm has a pronounced exponential rise with the evolving stream. Otherwise, our implementation has a slight increase in elapsed time, behaving more linearly as the stream evolves.

\figref{fig:Results2} (Cumulative Time) shows a clear advantage of using IS-TFIDF \& ICS, as the cumulative time rises at a slower pace when compared with the batch version. This is expectedly more pronounced as the stream evolves with time and more data arrives the system.

\figref{fig:Results2} (Speedup Ratio) shows that, although the IS-TFIDF implementation is slower in the first iterations, it starts being faster after some iterations.

%meter cumulative time chart

%meter speed-up chart

%meter numero de keywords obtidas

%meter numero de nos do grafo

%meter numero de palavras do corpus

\section{Conclusions}\label{Disc}

In this report, we present preliminary tests results for IS-TFIDF and ICS with the use of bipartite graphs. Although both tests were successful and positive regarding the benefits of IS-TFIDF and ICS, some further optimization of the code is possible. Therefore, there are further adjustments to be done, and it is expected that these improvements might increase the efficiency difference between R implementations of IS-TFIDF \& ICS and the batch implementations of TF-IDF and cosine similarity.

\section*{Acknowledgments}
This work was fully financed by the Faculty of Engineering of the Porto University. Rui Portocarrero Sarmento also gratefully acknowledges funding from FCT (Portuguese Foundation for Science and Technology) through a PhD grant (SFRH/BD/119108/2016). The authors want to thank also to the reviewers for the constructive reviews provided in the development of this publication.

\bibliographystyle{apalike}
\bibliography{Report}

\end{document}